\documentclass[a4paper,11pt]{article}
\pdfoutput=1 
\usepackage{jheppub} 
\usepackage[T1]{fontenc} 
\usepackage{placeins}
\usepackage{upgreek}
\usepackage{multirow}
\usepackage{multicol}
\usepackage{makecell}
\usepackage{amsmath}
\usepackage{mathtools}
\usepackage{float}
\usepackage{bm}
\usepackage{sectsty}
\usepackage{array}

\newcommand {\sNN}[1]{$\sqrt{s_{\rm NN}} = #1$}

\newcommand {\pp}{$p$+$p$}
\newcommand {\pt}{p_{\rm T}}
\newcommand {\nse}{n\sigma_{\rm e}}
\newcommand {\effPho}{\varepsilon_{\rm PHE}}

\newcommand {\jpsi}{J/\psi}
\newcommand {\dzero}{D^{\rm 0}}
\newcommand {\gev}{GeV/$c^2$}
\newcommand {\gevc}{GeV/$c$}
\newcommand {\raa}{$R_{\rm AA}$}

\newcommand {\ncoll}{$N_{\rm coll}$}
\newcommand {\npart}{$N_{\rm part}$}

\title{\boldmath Measurement of electrons from open heavy-flavor hadron decays in Au+Au collisions at \sNN{200} GeV with the STAR detector}
\author[1]{The STAR Collaboration,\note{Corresponding author.}}
\collaborationImg{\hspace*{370pt}\includegraphics[height=40pt]{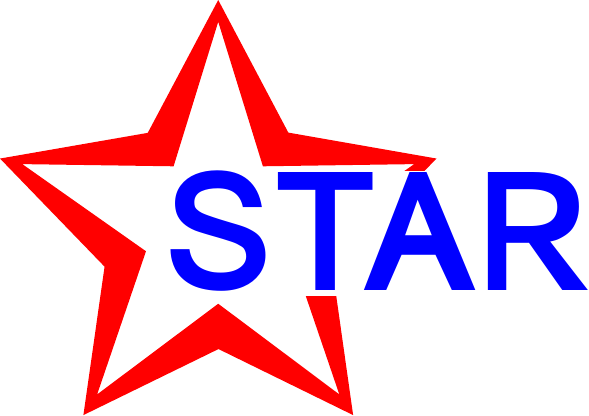}}
\collaboration{The STAR Collaboration\vspace*{-2em}}

\emailAdd{star-publication@bnl.gov}

\allsectionsfont{\normalfont\large\bfseries\boldmath}
\newcolumntype{L}[1]{>{\raggedright\let\newline\\\arraybackslash\hspace{0pt}}m{#1}}
\newcolumntype{C}[1]{>{\centering\let\newline\\\arraybackslash\hspace{0pt}}m{#1}}
\newcolumntype{R}[1]{>{\raggedleft\let\newline\\\arraybackslash\hspace{0pt}}m{#1}}
\newcolumntype{P}[1]{>{\raggedright\arraybackslash}p{#1}}
\newcolumntype{T}[1]{>{\raggedleft\arraybackslash}p{#1}}

\abstract{We report a new measurement of the production of electrons from open heavy-flavor hadron decays (HFEs) at mid-rapidity ($|y|<$ 0.7) in Au+Au collisions at \sNN{200} GeV. Invariant yields of HFEs are measured for the transverse momentum range of $3.5 < \pt < 9$ \gevc\ in various configurations of the collision geometry. The HFE yields in head-on Au+Au collisions are suppressed by approximately a factor of 2 compared to that in \pp\ collisions scaled by the average number of binary collisions, indicating strong interactions between heavy quarks and the hot and dense medium created in heavy-ion collisions. Comparison of these results with models provides additional tests of theoretical calculations of heavy quark energy loss in the quark-gluon plasma.} 
\begin{document} 
\maketitle
\flushbottom
\section{\label{sec:intro}Introduction}
Ultra-relativistic heavy-ion collisions provide a unique opportunity for studying Quantum Chromodynamics (QCD) in laboratories. The force that binds quarks together in nucleons can be screened at sufficiently high energy density, leading to a transition from ordinary nuclear matter to a new phase called the Quark-Gluon Plasma (QGP), whose properties are governed by partonic degrees of freedom. This state of matter is hypothesized to have existed in the early universe, a few millionths of a second after the Big Bang~\cite{ref:QGP1, ref:QGP2}. Experiments at the Relativistic Heavy Ion Collider (RHIC) and Large Hadron Collider (LHC) have provided strong evidence that a strongly-interacting QGP is created in collisions of heavy ions at RHIC and the LHC~\cite{BRAHMS:white:paper,STAR:white:paper,PHENIX:white:paper,PHOBOS:white:paper,LHC,RL}.

Owing to their large masses, heavy quarks, including charm ($c$) and beauty ($b$) quarks, are produced predominantly via hard partonic scatterings at early stages of a heavy-ion collision, and the thermal production in the QGP is negligible~\cite{PhysRevC.51.2177}. They subsequently probe the entire evolution of the system created in the collision, including the partonic phase of the QGP, hadronization and the hadronic phase \cite{hfreview1, hfreview3}. In particular, heavy quarks lose energy through interactions with the QGP via both collisional and radiative processes, with the former dominating at relatively low transverse momentum ($\pt$) and the latter taking over at high $\pt$. These interactions modify the momentum distributions of heavy quarks in heavy-ion collisions compared to that in \pp\ collisions, and measurements of such modifications provide important insights into the properties of the QGP. Furthermore, beauty quarks are expected to lose less energy than charm quarks because of their larger mass \cite{Dokshitzer:2001zm,Elias:2014hua}, and therefore separate measurements of charm and beauty quarks will further contribute to our understanding of the QGP. Significant suppression of charm meson yields at large $\pt$ has been observed at both RHIC and the LHC ~\cite{D0_STAR, D0_ALICE, D0_ALICE1, D0_CMS}, suggesting substantial energy loss experienced by charm quarks during propagation through the QGP medium. At the LHC, yields of beauty mesons \cite{B0_CMS}, as well as $\jpsi$ \cite{BtoJpsi_ALICE,BtoJpsi_CMS} and $\dzero$ \cite{BtoD_CMS,BtoD_ALICE} from $b$-hadron decays, are found to be less suppressed than charm hadrons, consistent with the expected mass dependence of the parton energy loss.


Electrons\footnote{Unless specified otherwise, electrons referred to here include both electrons and positrons and results are presented as $\frac{e^++e^-}{2}$.} from semileptonic decays of heavy-flavor hadrons (HFEs) are also widely used for measuring heavy quark production in heavy-ion collisions \cite{STAR:NPE,PHENIX:NPE,ALICE:NPE1,ALICE:NPE2}. Although they provide weaker constraints on parent heavy quark kinematics than heavy-flavor hadrons, the semileptonic decays of heavy-flavor hadrons have larger branching ratios and dedicated electron triggers can be utilized to sample large luminosities, making them experimentally more accessible. The HFE sample is usually a mixture of electrons from both charm and beauty hadron decays, with the latter constituting more than half of the whole sample above 5 \gevc\ in \pp\ collisions at $\sqrt{s} = 200$ GeV \cite{STARppb, PHENIXppb}. It is the main channel for accessing beauty quark production at RHIC. The inclusive HFE production in Au+Au collisions at \sNN{200} GeV has been studied by the STAR~\cite{STAR:NPE} and PHENIX~\cite{PHENIX:NPE,PHENIX:NPE2} experiments. However, these results have large uncertainties at high $\pt$, where the beauty quark contribution is the largest, and the previous STAR measurement only focused on head-on collisions. This calls for comprehensive measurements of HFE yield modifications at high $\pt$ with improved precision at RHIC, which also provide essential inputs for deriving the yield suppression of electrons from charm and beauty hadron decays separately~\cite{STAR:NPE2}.

In this article, we report a new differential measurement of the HFE production within $3.5 < \pt < 9$ \gevc\ at mid-rapidity ($|y|<$ 0.7) across different centrality bins (0-10\%, 10-20\%, 20-40\%, and 40-80\%) in Au+Au collisions at \sNN{200} GeV, while the result for the 0-80\% centrality bin has been recently reported in~\cite{STAR:NPE2}. The paper is organized as follows. In Sec.~\ref{sec:experiment}, components of the STAR detector relevant to this analysis are briefly discussed. Section~\ref{sec:analysis} is dedicated to the details of the data analysis of HFE production. Finally, results are reported and compared to previously published results and model calculations in Sec.~\ref{sec:results}.
\section{\label{sec:experiment}Experiment and datasets}
This work uses Au+Au collisions at \sNN{200} GeV recorded by the STAR experiment in 2014, utilizing the high-energy triggers, {\it i.e.} High Tower (HT) triggers, in addition to the minimum bias trigger condition based on the Vertex Position Detectors (VPDs)~\cite{ref:vpd_det}. The minimum bias trigger is defined by requiring coincidence signals between the two VPDs, with each VPD covering approximately half of the solid angle within the pseudorapidity ($\eta$) range of $4.24<|\eta|< 5.1$ on each side of the collision region. The HT trigger requires at least one tower in the Barrel Electromagnetic Calorimeter (BEMC)~\cite{ref:bemc_det} above a transverse energy threshold ($E_{\rm T}$). Events selected by two HT triggers of different thresholds are used: HT1 with $E_{\rm T} > 3.5$ GeV and HT2 with $E_{\rm T} > 4.2$ GeV, corresponding to integrated luminosities of 1.0 and 5.2 $\rm nb^{-1}$, respectively. The location of the collision vertex along the beam pipe direction can be calculated based on the timing information from the VPDs ($V_z^{\rm VPD}$) and reconstructed based on charged particle trajectories in the Time Projection Chamber (TPC) ($V_z^{\rm TPC}$)~\cite{ref:tpc_det}. To remove pile-up events, the $V_z^{\rm TPC}$ is required to be consistent with $V_z^{\rm VPD}$ within 3 cm, {\it i.e.} $|V_z^{\rm TPC} - V_z^{\rm VPD}|<3$ cm. Furthermore, a cut of $|V_z^{\rm TPC}|<30\, \rm cm$ is applied to ensure uniform TPC acceptance.

Two main subdetectors, the TPC and the BEMC, are used to reconstruct charged tracks and perform Particle IDentification (PID). The TPC, covering full azimuth within $|\eta|<1$, provides tracking, momentum determination and PID via measuring ionization energy loss ($dE/dx$). The BEMC, covering $|\eta|<1$ and full azimuth, can trigger on, and identify high-$\pt$ electrons. The BEMC is also equipped with a Barrel Shower Maximum Detector (BSMD) at a depth of 5.6 radiation lengths, which measures the shape and position of electromagnetic showers in the BEMC to further enhance electron identification capability. The multiplicity of charged particles in the TPC within $|\eta|<0.5$ is compared with a Glauber model~\cite{Miller:2007} to determine the collision centrality~\cite{D0_STAR}. Central (peripheral) events refer to collisions where incoming nuclei overlap with each other the most (least).
\section{\label{sec:analysis}Analysis details}
Experimentally identified electron candidates, called inclusive electron (INE) candidates, consist primarily of four components:
\begin{itemize}
\item Electrons from open heavy-flavor hadron (including non-prompt $J/\psi$) decays
\item Hadron contamination
\item Photonic electrons (PHE):
\begin{itemize}
\item photon conversion in the detector material: $\gamma \rightarrow e^{+}e^{-}$
\item $\pi^{0}$ Dalitz decay: $\pi^{0} \rightarrow e^{+}e^{-}\gamma\;[\rm{B.R.} = (1.174 \pm 0.035)\%]$
\item $\eta$ Dalitz decay: $\eta \rightarrow e^{+}e^{-}\gamma\;[\rm{B.R.} = (0.69 \pm 0.04)\%]$
\end{itemize}
\item Hadron decayed electrons (HDE):
\begin{itemize}
\item Heavy quarkonia contribution (prompt $J/\psi$ and $\Upsilon$)
\item Di-electron decays of light vector mesons ($\rho, \omega$ and $\phi$)
\item Drell-Yan contribution
\item Kaon semileptonic decays ($K_{e3}$)
\end{itemize}
\end{itemize}
The HFE invariant yield can be calculated as:
\begin{eqnarray}
\begin{aligned}
{Y_{\rm HFE}}&=Y_{\rm NPE}-Y_{\rm HDE}\\
&=\frac{1}{N_{\rm evt}} \times \frac{1}{2\pi \pt d\pt dy} \times \frac{N_{\rm INE} \times P_{\rm e}-N_{\rm PHE}/\epsilon_{\rm PHE}}{\epsilon_{\rm total}}-Y_{\rm HDE},
\label{eq:NPEyield}
\end{aligned}
\end{eqnarray}
where $Y_{\rm NPE}$ is the invariant yield of non-photonic electrons (NPE), $Y_{\rm HDE}$ is the invariant yield of HDE, $N_{\rm INE}$ is the raw yield of INE candidates, $P_{\rm e}$ is the electron purity in the INE candidates, $N_{\rm PHE}$ is the raw yield of PHE candidates, $\epsilon_{\rm PHE}$ is the PHE identification efficiency, $\epsilon_{\rm total}$ is the overall efficiency for triggering, tracking and particle identification of electrons, $y$ is the electron rapidity, and $N_{\rm evt}$ is the total numbers of sampled events. Here, NPE refers to the inclusive electron sample with hadron contamination and photonic electrons subtracted. 
\subsection{\label{sec:purity}Electron identification and purity}
A track reconstructed in the TPC is selected only if its Distance of Closest Approach (DCA) to the collision vertex is less than 1.5 cm, in order to suppress particles produced at secondary vertices. The number of TPC space points, also called ``TPC hits'', used for track reconstruction should be 20 or more to ensure good track quality, and also be larger than 52\% of the maximum possible number of TPC hits ($\leq$ 45) along the track trajectory to avoid split tracks. For achieving good $dE/dx$ resolution, the number of TPC hits used for $dE/dx$ calculation is required to be at least 15. Finally, only tracks within $|\eta|<0.7$ and with at least one hit in the first three TPC padrows are retained in order to minimize photonic electron background from photon conversions in the beam pipe support structure and TPC gas, respectively.

Electron candidates are identified using $dE/dx$ measured in the TPC, the ratio of track momentum measured by the TPC over energy deposition of the most energetic tower in the matched BEMC cluster ($p/E$), and the shower shape measured by the BSMD. To eliminate the momentum dependence of the $dE/dx$ value and its resolution, a normalized quantity, $\nse = \frac{ln(dE/dx_{mea})-ln(dE/dx_{th})}{\sigma(ln(dE/dx))}$, is used, where $dE/dx_{mea}$ is the measured value, $dE/dx_{th}$ is the theoretical value for electrons based on the Bichsel formalism~\cite{Bichsel:2006cs}, and $\sigma(ln(dE/dx))$ is the resolution. Tracks with $0.3 < p/E < 1.5$ and $-1.5<\nse<3.0$ are selected. To further discriminate electrons against hadrons, electron candidates are required to fire at least two strips in both the $\phi$ and $\eta$ planes of the BSMD, and the distances from the projected TPC track position to the reconstructed BEMC cluster position in the $\phi$ and $\eta$ planes to be less than 0.015 rad and 3 cm, respectively. 

\begin{figure}[tb]
\centering
\includegraphics[width=.495\textwidth]{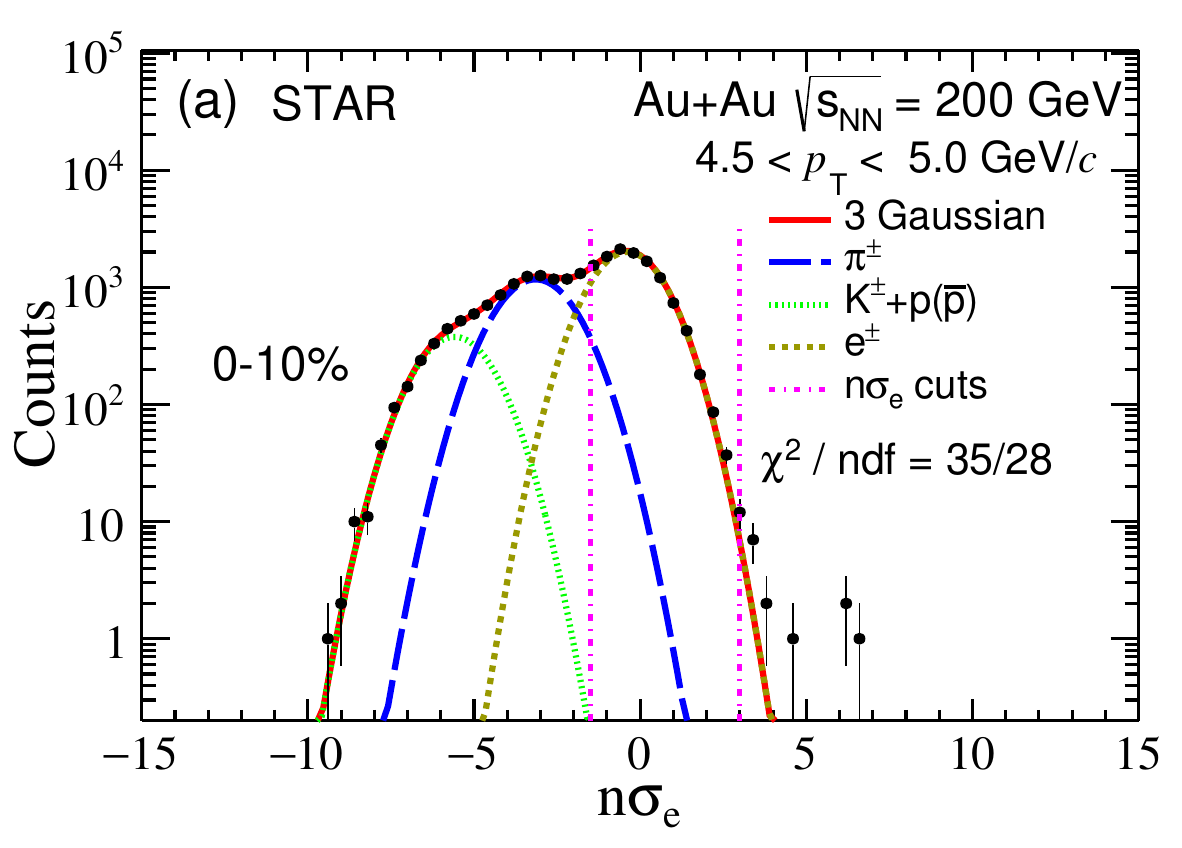}
\hfill
\includegraphics[width=.495\textwidth]{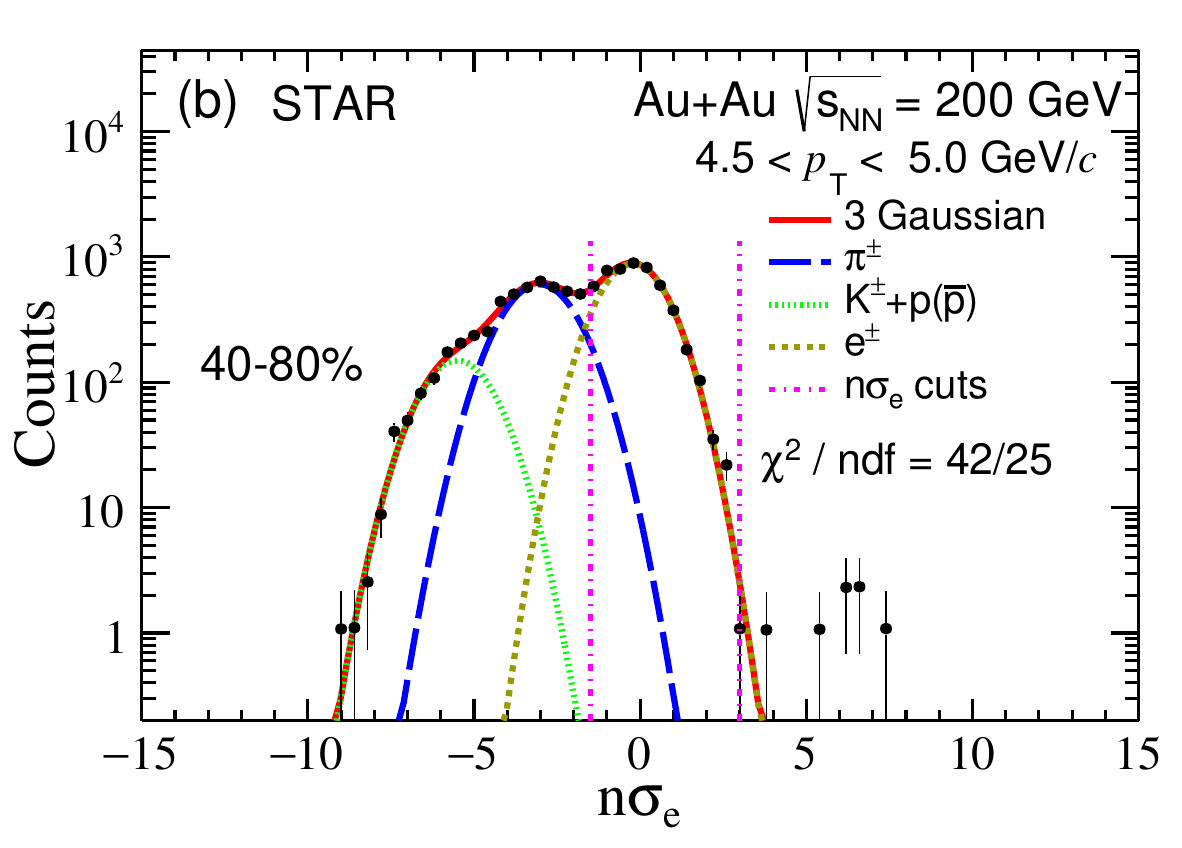}
\newline
\includegraphics[width=.495\textwidth]{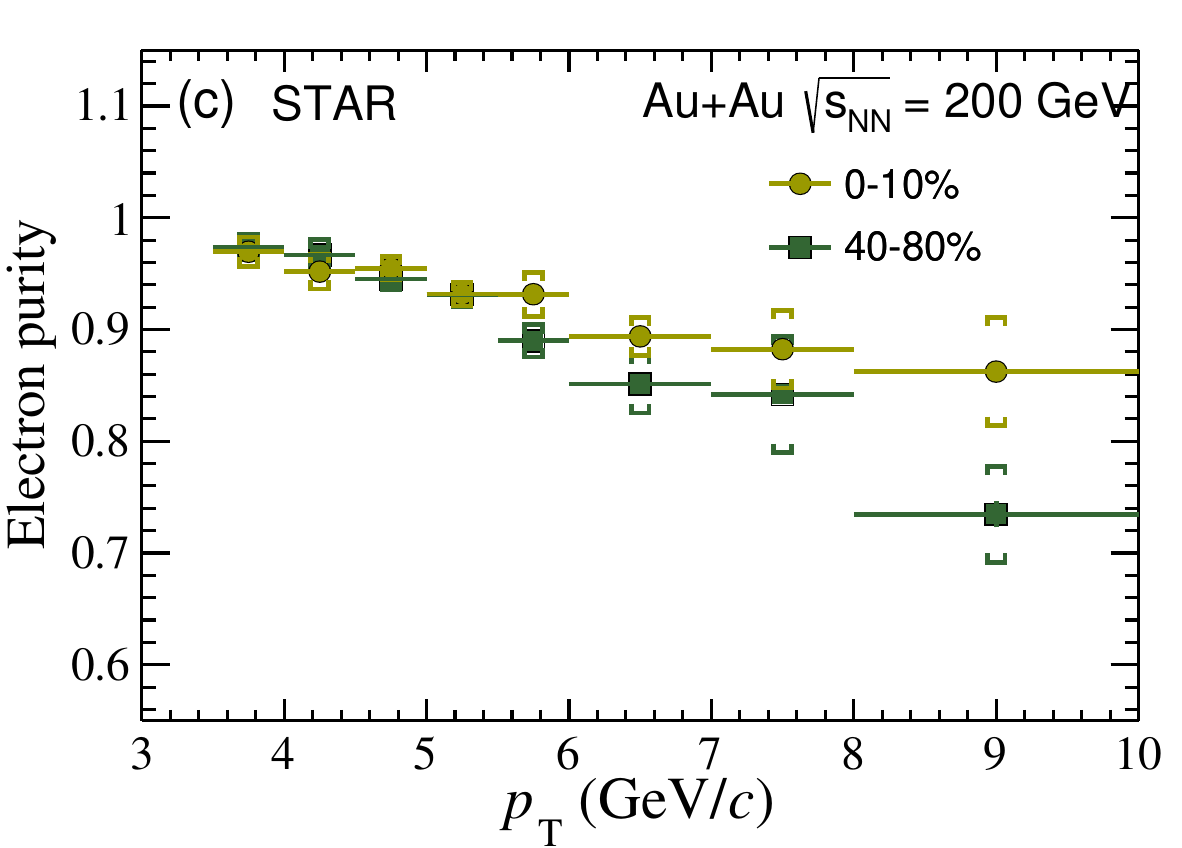}
\vspace*{-5pt}
\caption{\label{Fig:nseFits}(a) An example of $\nse$ distribution (black circles) with a three-Gaussian fit (solid red curve) for 4.5 $<\pt<$ 5.0 GeV/$c$ in 0-10\% central Au+Au collisions at \sNN{200} GeV. Gaussian functions (dotted curves in various colors) represent fits for different particle species. The dotted pink vertical lines indicate the $-1.5< \nse <3.0$ range used for electron selection. The small bump at 4 $< \nse <$ 10 is from track merging~\cite{STAR:dielectronau}. (b) Same as (a) except that it is for 40-80\% centrality. (c) Electron purity as a function of $\pt$ in 0-10\% central (yellow circles) and 40-80\% peripheral (green squares) Au+Au collisions. Vertical bars represent statistical uncertainties (smaller than the marker size) while boxes represent systematic uncertainties (details in Sec.~\ref{sec:syserr}). Horizontal bars indicate the bin width.} 
\end{figure}

TPC tracks that pass all the aforementioned cuts are classified as INE candidates. Figures \ref{Fig:nseFits} (a) and (b) show examples of $\nse$ distributions for 4.5 $<\pt<$ 5.0 \gevc\ in 0-10\% central and 40-80\% peripheral Au+Au collisions, respectively, for tracks satisfying all selection cuts except the $\nse$ cut. The integrals of $\nse$ distributions within $-1.5 <\nse< 3.0$ are the raw yields of INE candidates. To estimate the purity of the electron sample ($P_{\rm e}$) in the INE sample, a constrained fit to the $\nse$ distribution with three Gaussian functions representing $\pi^{\pm}$, $K^{\pm}$+$p$($\bar{p}$) and $e^{\pm}$, is performed and shown in Figs. \ref{Fig:nseFits} (a) and (b). For $\pi^{\pm}$ and $K^{\pm}$+$p$($\bar{p}$), initial mean $\nse$ values in the fit function are obtained from the Bichsel formalism~\cite{Bichsel:2006cs}, while initial widths are set to be 1. The mean and width of the Gaussian function for electrons are fixed according to the $\nse$ distribution of a pure electron sample consisting of photonic electrons (as described in Sec.~\ref{sec:pho:ele:id}) selected with an invariant mass cut of $M_{e^+e^-}< 0.1$ \gev. A good agreement between data and the fit function is seen, as evidenced by the $\chi^{2}/\rm ndf$ values shown in Figs. \ref{Fig:nseFits} (a) and (b). The electron purity is extracted by taking the ratio of the integral of the electron fit function to that of the overall fit function in the $\nse$ cut range ($-1.5 <\nse< 3.0$). The resulting purities as a function of electron $\pt$ in 0-10\% central and 40-80\% peripheral Au+Au collisions are shown in Fig. \ref{Fig:nseFits} (c). The purity decreases with increasing $\pt$ because the pion peak gets closer to the electron peak and the relative yield of pion to electron increases. At $\pt>$ 5.5 GeV/$c$, the purity seems smaller in 40-80\% peripheral collisions than that in 0-10\% central collisions, which is caused by the larger relative pion to electron yield in peripheral collisions.
\subsection{\label{sec:pho:ele:id}Photonic electron subtraction}
\begin{figure}[tb]
\centering
\includegraphics[width=.495\textwidth]{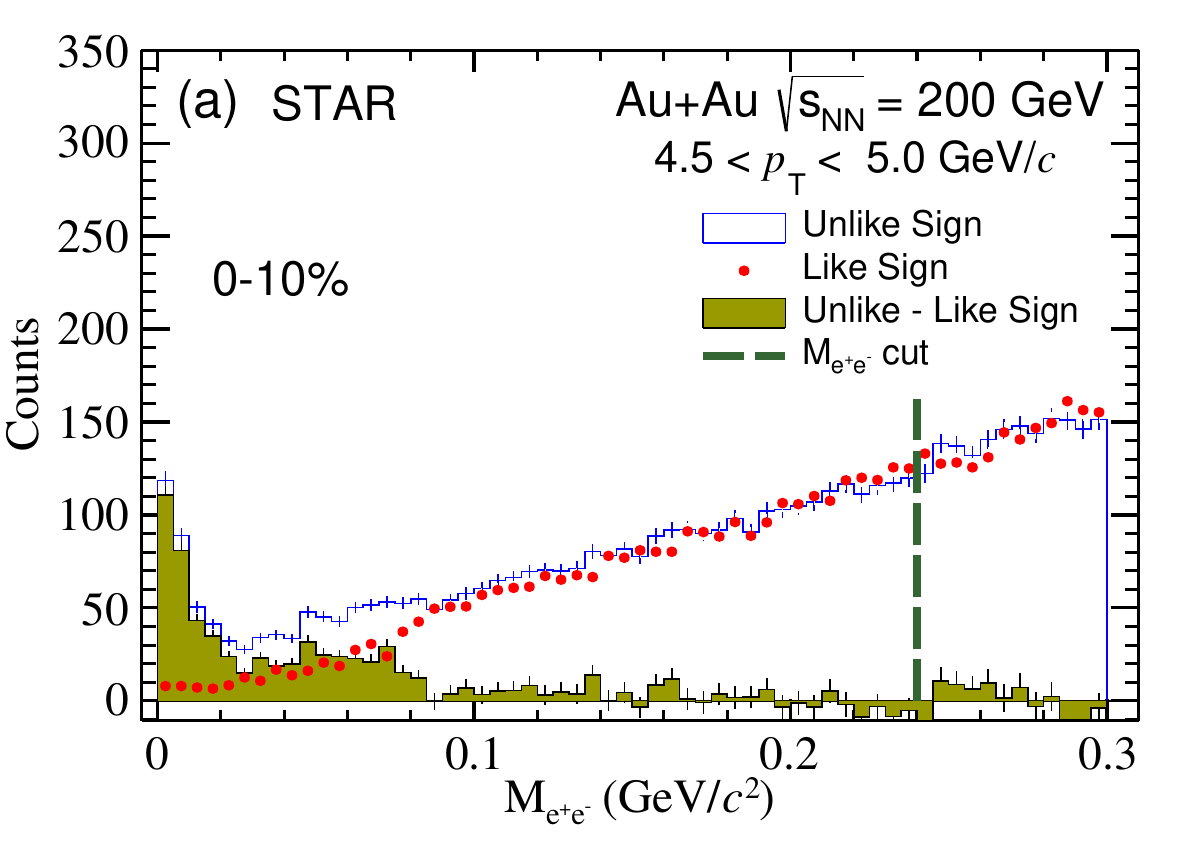}
\hfill
\includegraphics[width=.495\textwidth]{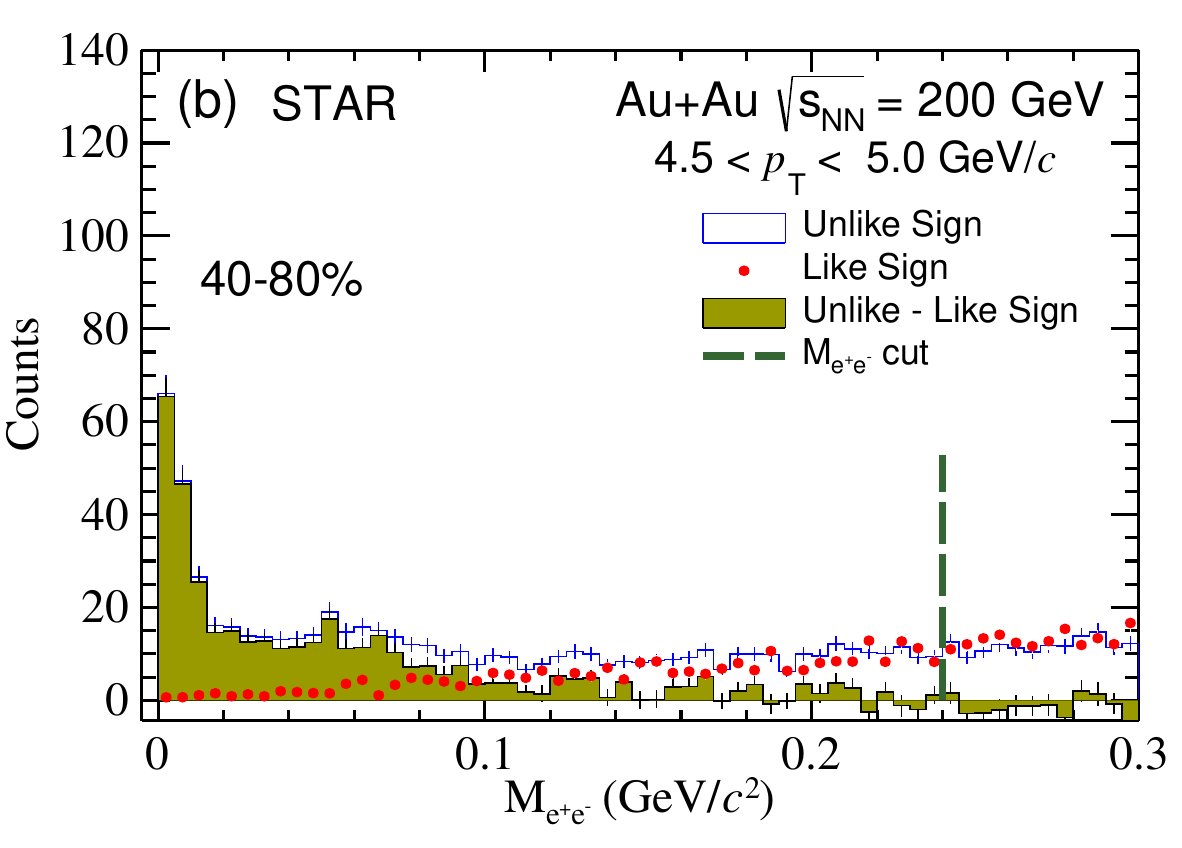}
\newline
\includegraphics[width=.495\textwidth]{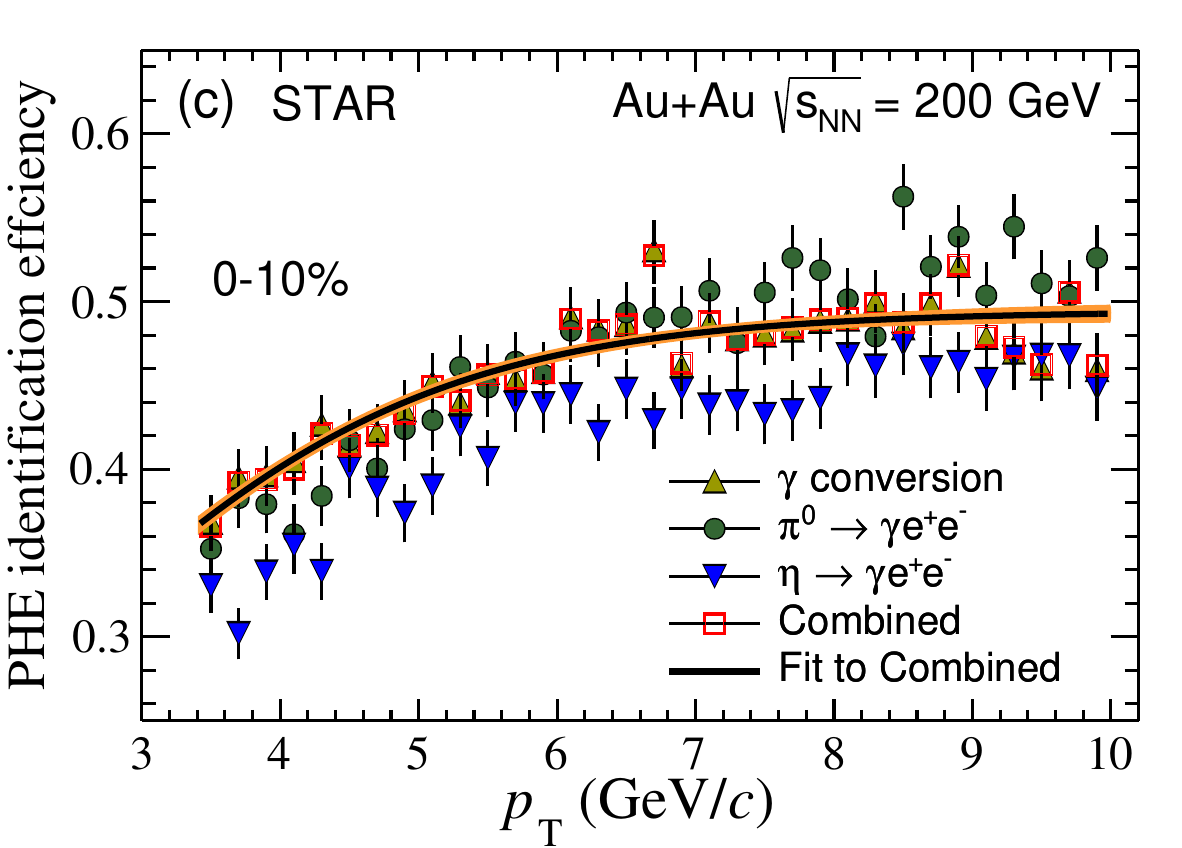}
\hfill
\includegraphics[width=.495\textwidth]{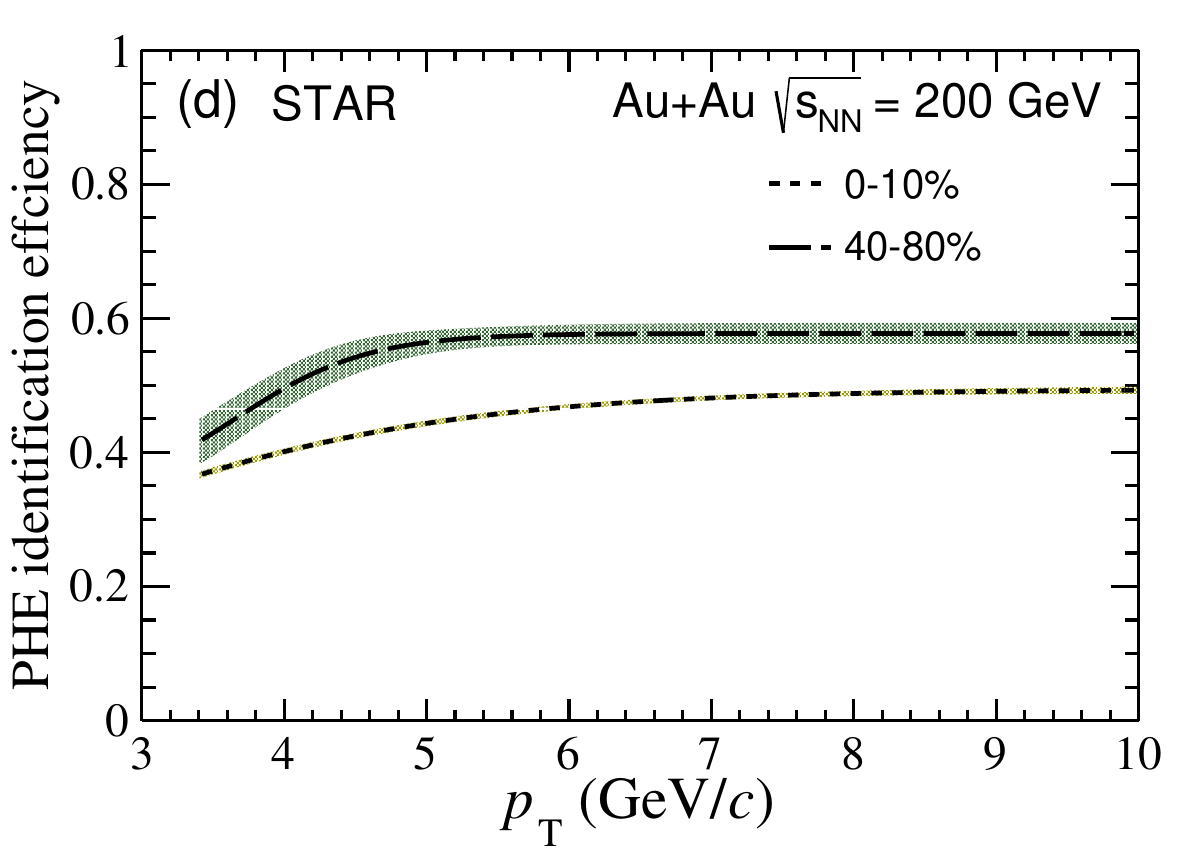}
\vspace*{-23pt}
\caption{\label{Fig:PhotoEleReco}(a) An example of invariant mass distributions for tagged electrons of $4.5<\pt<$ 5.0 \gevc\ in 0-10\% central Au+Au collisions at \sNN{200} GeV. The blue histogram labeled ``Unlike Sign'' shows the $e^+e^-$ pairs, the red circles labeled ``Like Sign'' mimic the combinatorial background, and the difference of the two labeled ``Unlike-Like Sign'' represents PHEs and is shown as the yellow histogram. The dotted green vertical line indicates the PHE selection cut. (b) Same as (a) except it is for 40-80\% centrality. (c) Combined PHE identification efficiency (red squares), together with a fit (black curve) and fit uncertainty (orange band), as a function of $\pt$ in 0-10\% central Au+Au collisions. PHE identification efficiencies for individual sources: photon conversion (yellow up triangles), $\pi^{0} $ Dalitz decay (green circles), and $\eta$ Dalitz decay (blue down triangles) are also shown. (d) Parametrizations of combined PHE identification efficiencies in 0-10\% central (dotted line) and 40-80\% peripheral (long dashed line) Au+Au collisions, with the uncertainties drawn as bands.}
\end{figure}
There are primarily two sources of PHEs: photon conversion and Dalitz decays of $\pi^{0}$ and $\eta$ mesons. Among the INE candidates, PHEs are found by paring them (tagged electrons) with oppositely-charged tracks (partner electrons) reconstructed in the TPC, denoted as unlike-sign pairs~\cite{STAR:ppNPE}. Tagged electrons are also paired with tracks of the same charge to construct like-sign distributions from a sum of $e^{+}e^{+}$ and $e^{-}e^{-}$ pairs, as estimates of misidentified PHEs arising from combinatorial background. Raw yields of PHEs are extracted by subtracting the invariant mass spectra of like-sign electron pairs from the unlike-sign ones, and applying an invariant mass cut of $M_{e^+e^-}< 0.24$ \gev, which takes into account the broadening of the invariant mass distribution with increasing tagged-electron $\pt$. Partner electrons are required to have $|\eta|< 1$, at least 15 TPC hits used for reconstruction, the ratio of the number of used to the maximum possible number of TPC hits larger than 0.52 and $\pt> 0.3$ \gevc. These requirements are less strict than those for tagged electrons in order to enhance the probability of finding PHEs. In addition, a maximum DCA of 1.0 cm between the two electron tracks is applied to ensure that the partner electron originates from the same production vertex as the tagged electron. Figures~\ref{Fig:PhotoEleReco} (a) and (b) show examples of invariant mass distributions for unlike-sign pairs, like-sign pairs, as well as differences between unlike- and like-sign pairs, for tagged electrons of $4.5 <\pt< 5.0$ \gevc\ in 0-10\% central and 40-80\% peripheral Au+Au collisions, respectively. The like-sign distributions are seen to match well unlike-sign distributions at $M_{e^+e^-}> 0.24$ \gev, where combinatorial background dominates. 

The PHE identification efficiency, $\effPho$, which accounts for finding a partner electron and passing the pair DCA and invariant mass cuts, is evaluated by embedding full GEANT~\cite{ref:geant} simulations of $\gamma$, $\pi^0$ and $\eta$ decays in the STAR detector into real events, which then go through the same reconstruction and analysis software chain as real data. The decay processes are simulated with {\sc pythia 6.419}~\cite{ref:pythia}. Input $\pi^0$ $\pt$ spectra in different centrality classes are taken as the average of charged and neutral pion spectra in 200 GeV Au+Au collisions measured by STAR and PHENIX experiments~\cite{STAR:pi0, PHENIX:pi01, PHENIX:pi02}, while the input $\pt$ spectra for $\eta$ are obtained from $\pi^0$ spectra assuming traverse mass ($m_{\rm T}$) scaling, {\it i.e.} replacing $\pt$ in the $\pi^0$ spectra by $\sqrt{\pt^2 - m_{\rm \pi}^2 + m_{\rm \eta}^2}$. The input rapidity distributions of $\pi^0$ and $\eta$ are parametrized  with a Gaussian-like function $\cosh^{-2}\left( \frac{3y}{4\sigma(1-y^2/(2\sqrt{s}/m))}\right)$, where $\sigma=\sqrt{\ln(\sqrt{s}/(2m_{\rm N}))}$, $\sqrt{s}$ is a nucleon-nucleon center of mass energy, $m$ is the particle mass, $y$ is the particle rapidity, and $m_{\rm N}$ is the nucleon mass~\cite{ref:INP, STAR:dielectron, CERES:dielectron}. On the other hand, input spectra for photons are a combination of direct photon spectra measured by the STAR experiment~\cite{STAR:photon} and decayed photon spectra from $\pi^0 \rightarrow \gamma \gamma$/$e^+e^-\gamma$ and $\eta \rightarrow \gamma \gamma$/$e^+e^-\gamma$ processes obtained using the aforementioned $\pi^0$ and $\eta$ spectra for the Dalitz decay as inputs to {\sc pythia}. Figure~\ref{Fig:PhotoEleReco} (c) shows the combined PHE identification efficiency from photon conversion and Dalitz decays as a function of $\pt$ in 0-10\% central Au+Au collisions, along with a fit using the functional form $A/(e^{-(p_{\rm T}-p_{\rm 0})/p_{\rm 1}}+1)+C$, where $A$, $p_{\rm 0}$, $p_{\rm 1}$, and $C$ are free parameters. The individual $\effPho$ distributions for $\gamma$ conversion and two types of Dalitz decays are also shown in Fig.~\ref{Fig:PhotoEleReco} (c). Figure \ref{Fig:PhotoEleReco} (d) shows fits to combined $\effPho$ as a function of $\pt$ in 0-10\% central and 40-80\% peripheral Au+Au collisions. As expected, $\effPho$ is lower in central collisions than in peripheral collisions due to the decreasing tracking efficiency for partner electrons with increasing TPC occupancy in central collisions.

\begin{figure}[tb]
\centering
\includegraphics[width=.495\textwidth]{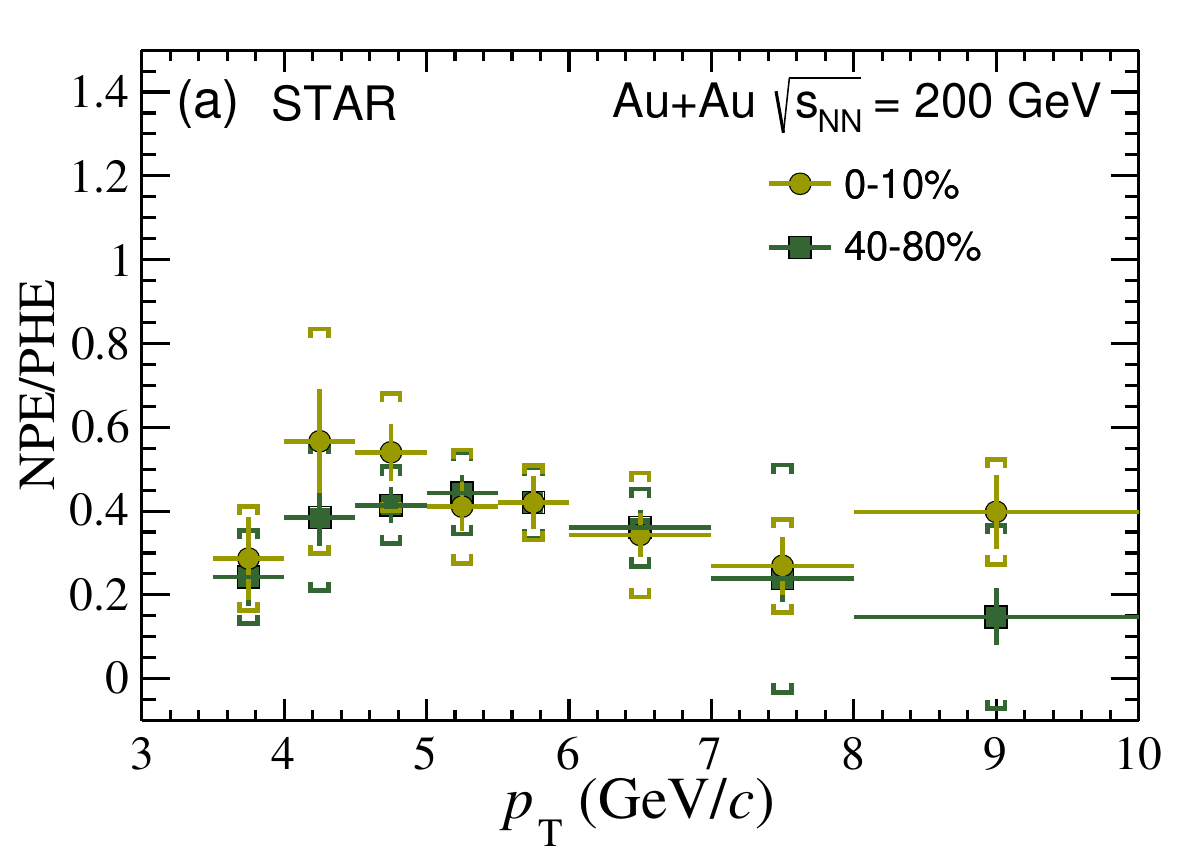}
\hfill
\includegraphics[width=.495\textwidth]{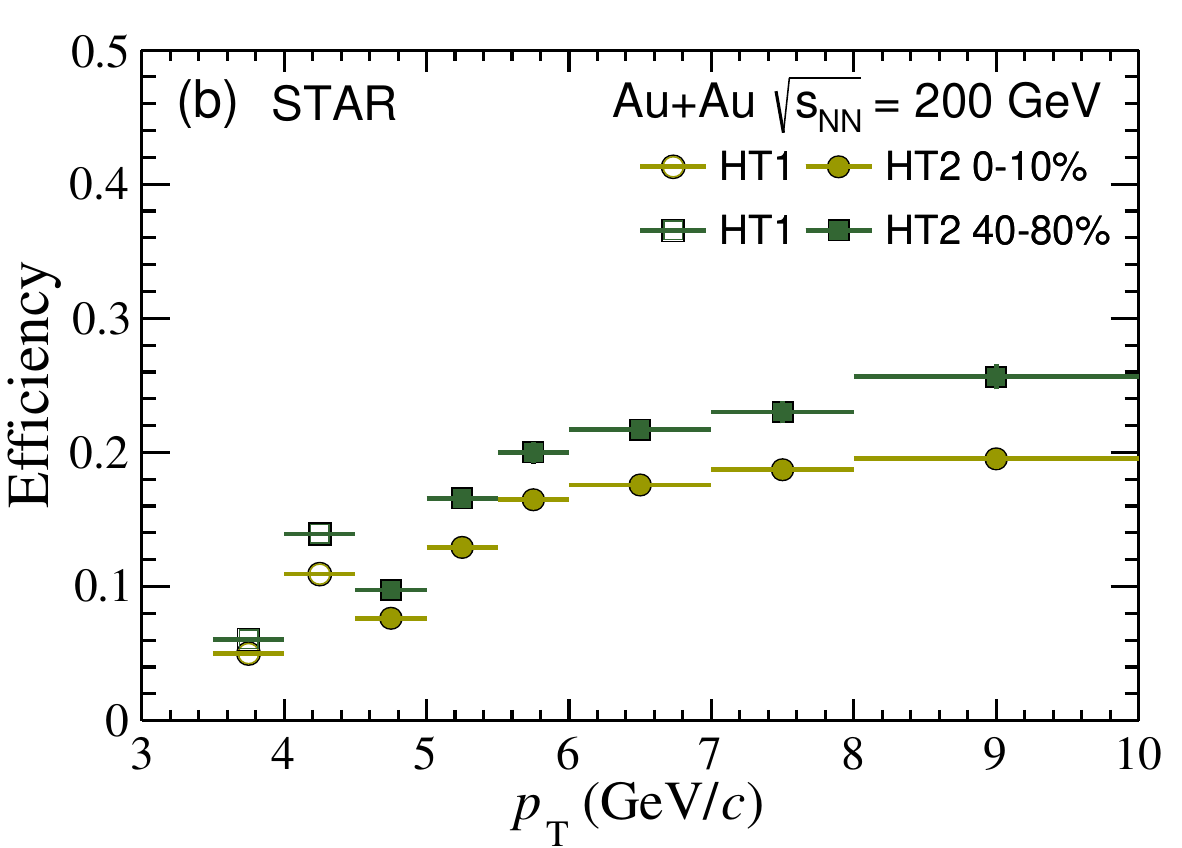}
\vspace*{-23pt}
\caption{\label{Fig:StoB_ratio_Efficiency}(a) Ratios of NPE to PHE as a function of $\pt$ in 0-10\% central (yellow circles) and 40-80\% peripheral (green squares) Au+Au collisions at \sNN{200} GeV. Vertical bars represent statistical uncertainties while boxes represent systematic uncertainties (details in Sec.~\ref{sec:syserr}). Horizontal bars indicate the bin width. (b) Overall electron detection efficiency [$\epsilon_{\rm total}$ in Eq.~(\ref{eq:NPEyield})] as a function of $\pt$ in 0-10\% central (yellow circles) and 40-80\% peripheral (green squares) Au+Au collisions. Open and solid points are the efficiencies for HT1- and HT2-triggered electrons, respectively. Vertical bars represent uncertainties, which are smaller than the marker size in many cases. Horizontal bars indicate the bin width.}
\end{figure}

The raw NPE yields can be obtained by statistically subtracting hadron contamination and efficiency-corrected PHE yields from INE candidates. Figure~\ref{Fig:StoB_ratio_Efficiency} (a) shows the yield ratios of NPE [$N_{\rm INE}\times P_{\rm e} -  N_{\rm PHE}/\effPho$ in Eq.~(\ref{eq:NPEyield})] to PHE background [$N_{\rm PHE}/\effPho$ in Eq.~(\ref{eq:NPEyield})] as a function of $\pt$ in 0-10\% central and 40-80\% peripheral Au+Au collisions, which are seen to be similar. These ratios are smaller than those in the previous STAR analysis based on 200 GeV $p$+$p$ collisions recorded in 2012~\cite{STAR:ppNPE}, due to the added material of the heavy flavor tracker~\cite{hft} and its support structure installed in 2014. 

\subsection{\label{sec:eff}Efficiency correction}
The NPE yields are obtained by correcting raw NPE yields for the overall efficiency [$\epsilon_{\rm total}$ in Eq.~(\ref{eq:NPEyield})]. The $\epsilon_{\rm total}$ is evaluated using the same approach as in Ref.~\cite{STAR:ppNPE}, which is briefly summarized here. The detector acceptance and efficiencies of TPC tracking, BEMC electron identification, and HT triggering are estimated by embedding single electrons into real data. The electron identification efficiencies of the TPC $\nse$ and BSMD requirements are evaluated using a data-driven method, \textit{i.e.}, taking the ratio of electrons with and without the $\nse$ or BSMD selection in the pure electron sample. Figure \ref{Fig:StoB_ratio_Efficiency} (b) shows the overall efficiencies as a function of $\pt$ for HT1- and HT2-triggered electrons in 0-10\% central and 40-80\% peripheral Au+Au collisions. The higher efficiency in peripheral collisions than central collisions is again due to the reduced TPC occupancy. The increasing efficiency with $\pt$ for HT1 and HT2 trigger, and the efficiency dropping from HT1 to HT2 trigger are mainly driven by the HT trigger threshold.

\subsection{\label{sec:hde}Hadron decayed electron background}
There are four sources for HDEs, including quarkonia, light vector mesons, Drell-Yan and Kaon semileptonic decays, as mentioned at the beginning of this section. 

\begin{figure}[tb]
\centering
\includegraphics[width=.495\textwidth]{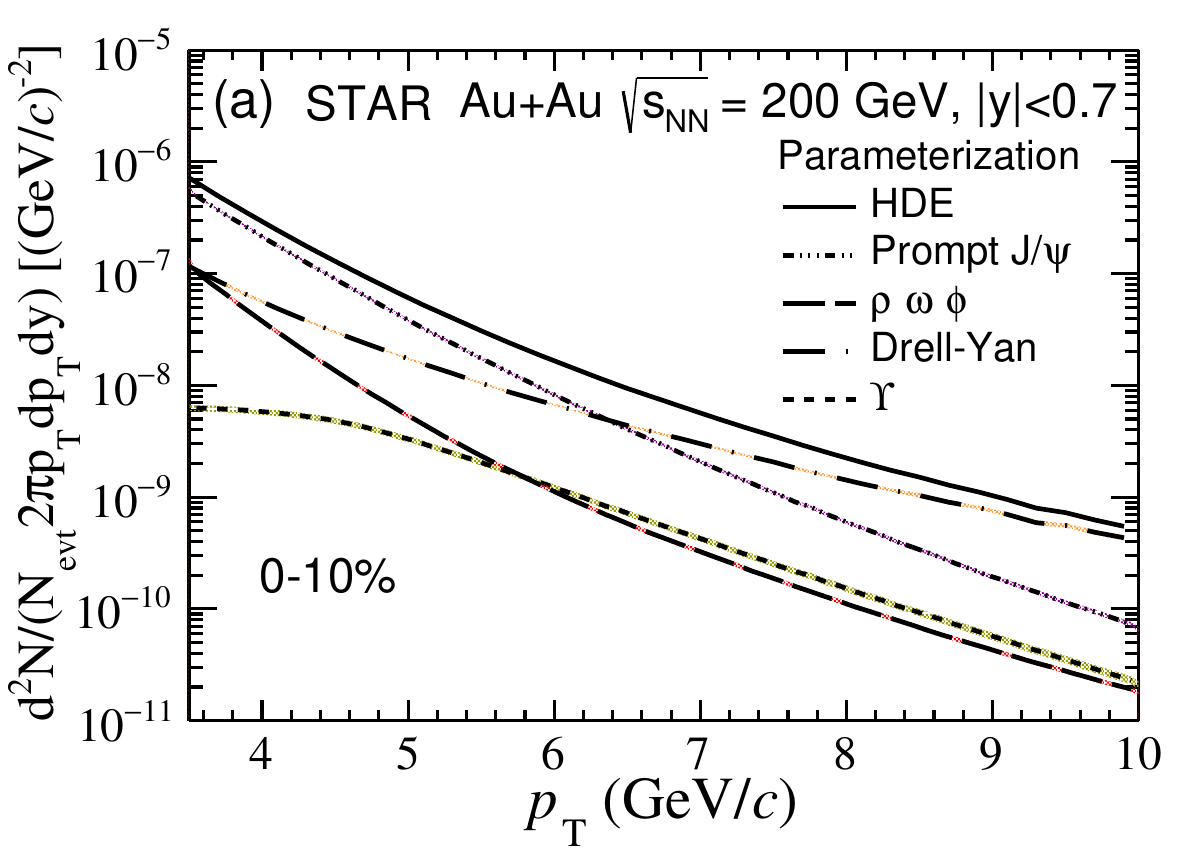}
\hfill
\includegraphics[width=.495\textwidth]{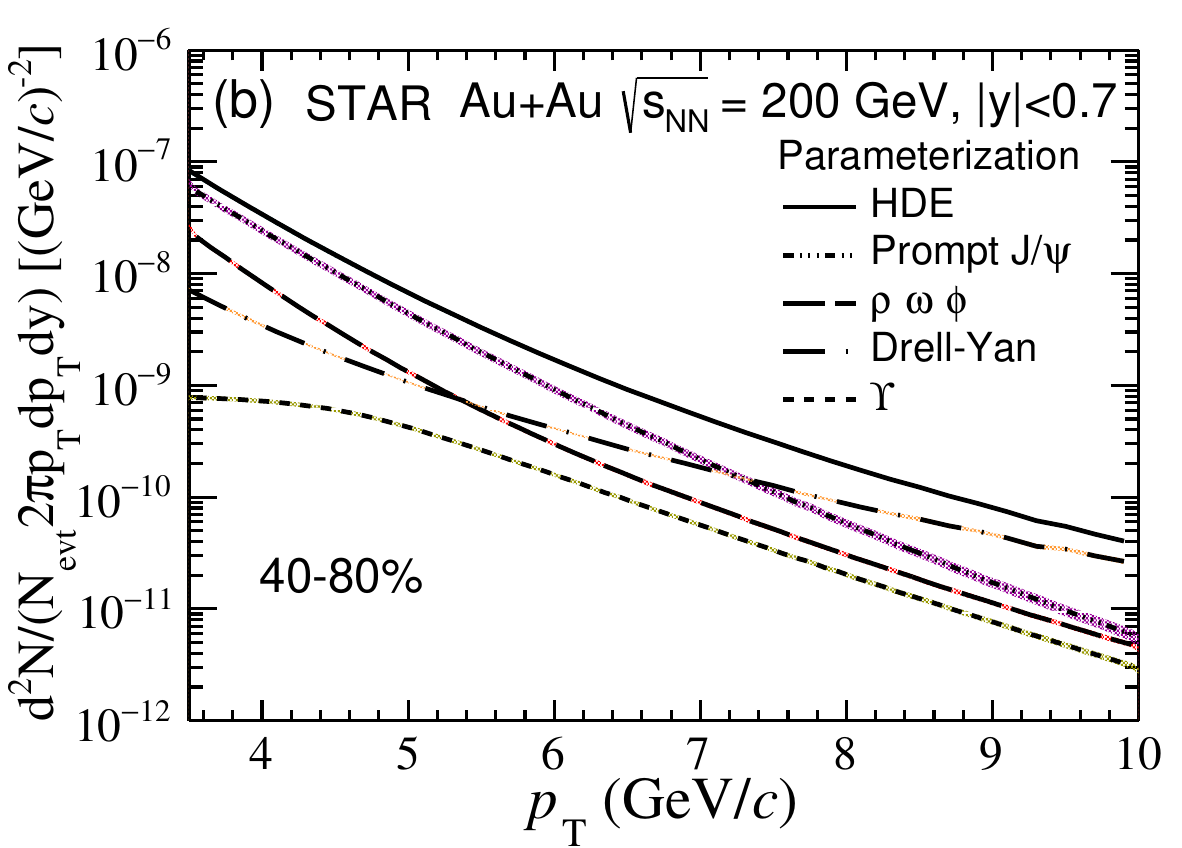}
\vspace*{-23pt}
\caption{\label{Fig:HDE}Invariant yields of electrons from decays of prompt $\jpsi$ (dot-dashed line), $\Upsilon$ (dotted line), Drell-Yan (long dash-dotted line), light vector mesons (long dashed line) and the combined HDE contribution (solid line), estimated utilizing experimental measurements, theoretical calculations, and {\sc pythia} and {\sc EvtGen} event generators, in 0-10\% central (a) and 40-80\% peripheral (b) Au+Au collisions at \sNN{200} GeV. Color bands represent systematic uncertainties. See text for details.} 
\end{figure}

The {\sc EvtGen} event generator~\cite{ref:evtgen} is used to decay prompt $\jpsi$ to electrons. The input $\pt$ spectra for prompt $\jpsi$ production are obtained from the published inclusive $\jpsi$ measurements~\cite{STAR_jpsi} parametrized with the Tsallis statistics~\cite{ref:TS, ref:TS1, ref:TS2} and with the non-prompt $\jpsi$ contribution subtracted based on Fixed Order plus Next-to-Leading Logarithms (FONLL) calculation~\cite{ref:QCDCB} plus Color Evaporation Model (CEM) ~\cite{ref:fcem1, ref:fcem2}. The rapidity distribution of prompt $\jpsi$ is taken from {\sc pythia}. The resulting invariant yields of decayed electrons in 0-10\% central and 40-80\% peripheral Au+Au collisions are represented by dot-dashed lines in Fig.~\ref{Fig:HDE}. For the $\Upsilon$ contribution, a model calculation~\cite{ref:upsilonmodel} indicates no significant $\pt$ dependence of $\Upsilon$ suppression in Au+Au collisions at \sNN{200} GeV, which is consistent with STAR measurements within uncertainties~\cite{ref:upsilon}. Therefore, the $\Upsilon$ decayed electrons in Au+Au collisions are estimated by scaling up their yield in 200 GeV $p$+$p$ collisions~\cite{STAR:ppNPE} by the average number of binary collisions (\ncoll)~\cite{D0_STAR}, incorporating model predictions of $\Upsilon$ suppression in the QGP~\cite{ref:upsilonmodel}. Invariant yields of electrons from $\Upsilon$ decays are shown as dotted lines in Fig.~\ref{Fig:HDE}. 

The $\pt$ spectra of light vector mesons, $\rho$, $\omega$, and $\phi$, in different centrality classes of Au+Au collisions are obtained by assuming $m_{\rm T}$ scaling based on the $\pi^0$ spectra in corresponding centrality classes, which are further scaled by the integrated yield ratio of light vector mesons over $\pi^0$ in 0-80\% centrality class~\cite{STAR:dielectronau}. Their rapidity distributions are obtained following the Gaussian-like function introduced in Sec.~\ref{sec:pho:ele:id}. {\sc pythia} is used to model the di-electron decay of the $\rho$ meson, while {\sc EvtGen} is used for $\omega$ and $\phi$. Invariant yields of resulting decayed electrons are illustrated as long dashed lines in Fig.~\ref{Fig:HDE} for 0-10\% central and 40-80\% peripheral Au+Au collisions.

For the Drell-Yan contribution, it is estimated as the Drell-Yan $\rightarrow e$ yield from {\sc pythia} simulation of 200 GeV $p$+$p$ collisions~\cite{STAR:ppNPE} scaled by \ncoll\ assuming no cold or hot nuclear matter effects, and shown as long dash-dotted lines in Fig.~\ref{Fig:HDE}. Furthermore, simulation studies based on STAR acceptance have shown that the $K_{\rm e3}$ contribute less than 2\% to HDE for $\pt>$ 3 \gevc\ in Au+Au collisions at \sNN{200} GeV~\cite{ref:ke3}, and are thus neglected. 

The overall $\rm HDE$ contributions in 0-10\% central and 40-80\% peripheral collisions, represented by solid lines in Fig.~\ref{Fig:HDE}, are subtracted from the NPE sample, and the remaining HFE yields are reported in Sec.~\ref{sec:results}. These contributions amount to a $\sim$15\%, $\sim$16\%, $\sim$18\% and $\sim$19\% reduction to the NPE yield in the measured $\pt$ region for 0-10\%, 10-20\%, 20-40\% and 40-80\% collisions, respectively.
\subsection{\label{sec:syserr}Systematic uncertainties}
For the NPE reconstruction efficiency, the uncertainties are estimated partially by changing the track quality and PID cuts in data and simulation simultaneously and checking variations in the corrected NPE yield. These include: (i) the number of TPC hits used for track reconstruction ($dE/dx$ calculation) from 20 (15) to 25 (18), and the larger variation of the two is taken; (ii) DCA from 1.5 cm to 1.0 cm; and (iii) $0.3 <p/E< 1.5$ to 0.6 $<p/E<$ 1.5 and $0.3 <p/E< 1.8$. Uncertainty in the HT trigger efficiency is evaluated by adjusting the trigger threshold in simulation by $\pm$ 5\%, originating from the uncertainties of the BEMC energy scale calibration. For the PID efficiency arising from BSMD requirements, its uncertainties are taken as the statistical errors of the pure electron sample in data used for estimating such an efficiency. The uncertainty of the $\nse$ cut efficiency is estimated from the parameter errors in fitting the $\nse$ distribution of the pure electron sample with a Gaussian function, taking into account the correlation between the mean and width parameters, and from varying the selection cut from $-1.5<\nse<3.0$ to $-1.0<\nse<3.0$. The uncertainties in electron purity are similarly estimated based on the uncertainties in the mean and width of Gaussian fits to the pure electron $\nse$ distributions. 

The PHE identification efficiency uncertainty stems from the uncertainties in simulation statistics, parametrizations of $\pi^0$ and $\eta$ spectra, branching ratios of electrons from $\pi^0$ and $\eta$ decays, tracking efficiency of partner electrons and variations in the PHE selection criteria, \textit{i.e.}, changing  maximum $M_{e^+e^-}$ from 0.24 \gev\ to 0.15 \gev\ and minimum partner electron $\pt$ from 0.3 GeV/$c$ to 0.2 GeV/$c$. The parametrization uncertainty is taken as the 68\% confidence interval of the fit function. Such an approach is also used in estimating the uncertainties in spectrum parametrization as described in the following.  

The uncertainty in estimating the HDE contribution includes those from $\jpsi$, $\Upsilon$, light vector meson, and Drell-Yan contributions. Uncertainties from parametrizating the inclusive $\jpsi$ spectrum and from FONLL+CEM calculations of the non-prompt $\jpsi$ contribution are taken into account. For the $\Upsilon$ contribution, uncertainties arise from measurements of $\Upsilon$ yields in $p$+$p$ collisions~\cite{STAR:ppNPE} and model calculations~\cite{ref:upsilonmodel}. Parametrization uncertainties of the $\pi^0$ spectra~\cite{STAR:pi0, PHENIX:pi01, PHENIX:pi02} as well as uncertainties in the measured yield ratios of light vector mesons to $\pi^0$~\cite{STAR:dielectronau} are also propagated to the decayed electron invariant yields. Finally, the uncertainty in the Drell-Yan contribution is from that of the results in $p$+$p$ collisions~\cite{STAR:ppNPE}. 

The total systematic uncertainty is obtained as the square root of the quadratic sum of individual sources. Table \ref{tab:syserror} summarizes the uncertainties from different sources and total uncertainties for HFE invariant yield measurements in different centrality intervals (0-10\%, 10-20\%, 20-40\%, 40-80\%). Global uncertainties, referred to in the following section, include those from the non-single diffractive cross section of $p$+$p$ collisions~\cite{STAR:D0} and \ncoll~\cite{D0_STAR}.
\begin{table}
\centering
\caption{\label{tab:syserror}Summary of individual and total systematic uncertainties, in percentage, for the $\rm HFE$ invariant yields in different centrality intervals (0-10\%, 10-20\%, 20-40\%, 40-80\%). The uncertainty ranges indicate variations with $\rm HFE$ $\pt$. In general, the uncertainty increases from low to high $\pt$.}
\begin{tabular}{c|cccc} \hline \hline
Source & \multicolumn{4}{c}{Systematic Uncertainty}\\ 
    &
       \hspace{0.5cm}0--10\% &
        \hspace{0.5cm}10--20\% &
        \hspace{0.5cm}20--40\% &
       \hspace{0.5cm}40--80\%\hspace{0.5cm}  \\ \hline \hline
$\rm NPE$ reconstruction efficiency & \hspace{0.5cm}9-27\% & \hspace{0.5cm}7-26\% & \hspace{0.5cm}5-23\% & 9-29\%\\
$\nse$ cut efficiency & \hspace{0.5cm}1-23\% & \hspace{0.5cm}1-6\% & \hspace{0.5cm}1-8\% & 1-7\%\\
Electron purity extraction & \hspace{0.5cm}4-23\% & \hspace{0.5cm}4-28\% &\hspace{0.5cm} 3-79\% & 4-76\%\\
$\rm PHE$ identification efficiency & \hspace{0.5cm}13-24\% & \hspace{0.5cm}13-29\% & \hspace{0.5cm}16-38\% & 15-70\%\\
$\rm HDE$ contribution & \hspace{0.5cm}1-2\% & \hspace{0.5cm}1-2\% & \hspace{0.5cm}1-3\% & 2-7\%\\
Total & \hspace{0.5cm}18-36\% & \hspace{0.5cm}17-37\% & \hspace{0.5cm}19-87\% & 19-107\%\\ \hline \hline
\end{tabular}
\end{table}
\begin{figure}[tb]
\centering
\includegraphics[width=.7\textwidth,page=1]{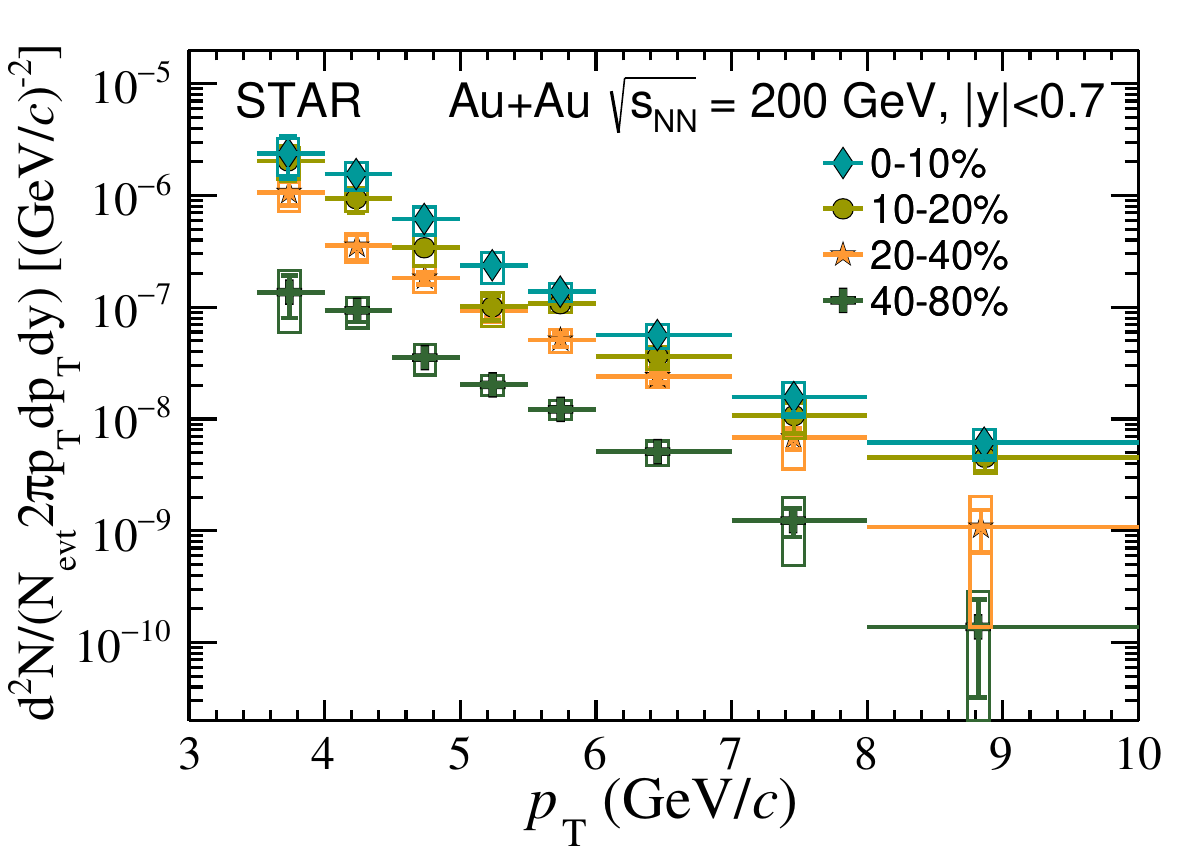}\vspace*{-10pt}
\caption{\label{Fig:HFE}HFE invariant yields in different centrality intervals of Au+Au collisions at \sNN{200} GeV. The vertical bars and the boxes represent statistical and systematic uncertainties, respectively. The horizontal bars indicate the bin width.}\vspace*{-14pt}
\end{figure}
\section{\label{sec:results}Results}
Following Eq.~\ref{eq:NPEyield}, the obtained invariant yields of HFEs within $|y|<0.7$ are shown in Fig.~\ref{Fig:HFE} as a function of $\pt$ for different centrality intervals (0-10\%, 10-20\%, 20-40\%, 40-80\%) in Au+Au collisions at \sNN{200} GeV.

\begin{figure}
\centering
\includegraphics[width=\textwidth,page=1]{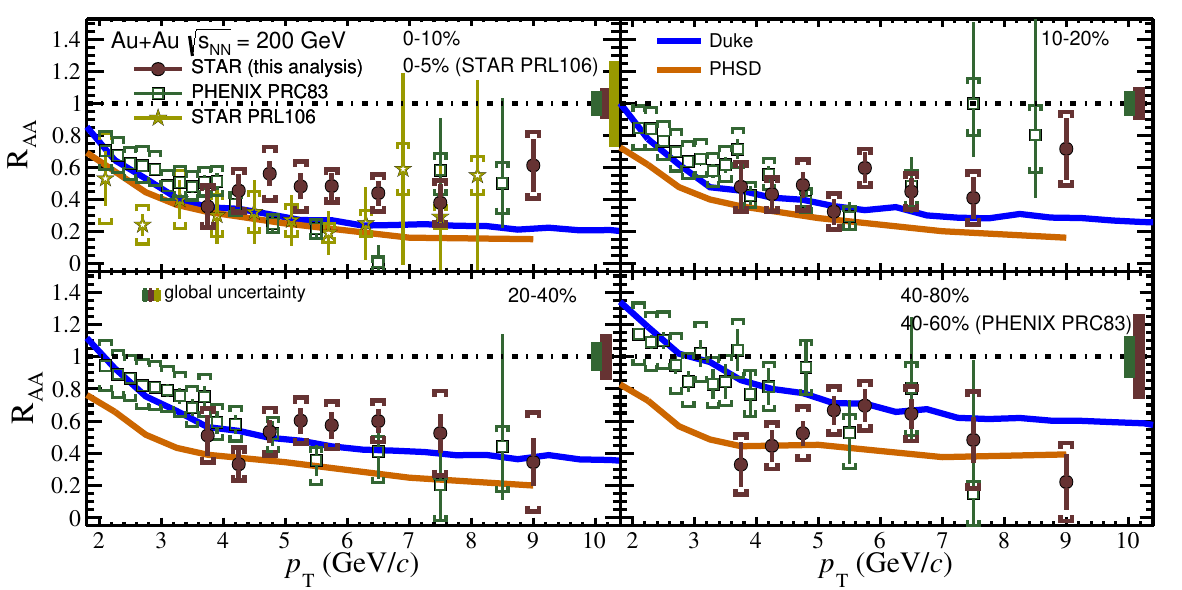}
\vspace*{-20pt}
\caption{\label{Fig:Raa}HFE \raa\ (red circles) as a function of $\pt$ in different centrality intervals of Au+Au collisions at \sNN{200} GeV, compared with STAR (yellow stars)~\cite{STAR:NPE} and PHENIX (green squares)~\cite{PHENIX:NPE} published results, and Duke (blue line)~\cite{ref:duke1} and PHSD (orange line)~\cite{ref:phsd1, ref:phsd2} model calculations. Vertical bars and boxes around data points represent combined statistical and systematic uncertainties from both Au+Au and \pp\ measurements, respectively. Boxes at unity show the global uncertainties, which for this analysis include the 8\% global uncertainty on $p$+$p$ reference ~\cite{STAR:D0} and the $N_{\rm coll}$ uncertainties. The left box is for PHENIX and the right one for STAR.}
\end{figure}

The nuclear modification factor (\raa) for HFEs is defined as: 
\begin{equation}
R_{\rm AA} =\frac{1}{N_{\rm coll}}\times \frac{\mathrm{d}N^2_{\rm AA}/(\mathrm{d}\pt \mathrm{d}y)}{\mathrm{d}N^2_{\rm pp}/(\mathrm{d}\pt \mathrm{d}y)},
\label{eq:RAA}
\end{equation}
where $\mathrm{d}N^2_{\rm AA}/(\mathrm{d}\pt \mathrm{d}y)$ and $\mathrm{d}N^2_{\rm pp}/(\mathrm{d}\pt \mathrm{d}y)$ are HFE yields in Au+Au and \pp\ collisions~\cite{STAR:ppNPE}, respectively. Figure~\ref{Fig:Raa} shows HFE $R_{\rm AA}$ as a function of $\pt$ in different centrality intervals of Au+Au collisions at \sNN{200} GeV. A suppression by about a factor of 2 is observed within $3.5<\pt<8.0$ \gevc\ in central and semi-central collisions, indicative of substantial energy loss of heavy quarks in the QGP. Within uncertainties, no significant $\pt$ dependence is observed in the measured $\pt$ range. Previous measurements by STAR~\cite{STAR:NPE} and PHENIX~\cite{PHENIX:NPE}, in which the STAR results include HDE contribution while the PHENIX results exclude both HDE and electrons from non-prompt $\jpsi$ decays, are also shown in Fig. \ref{Fig:Raa}. Compared to the PHENIX results~\cite{PHENIX:NPE}, precision of the new results is significantly improved for $\pt >6$ \gevc, while compared to previous STAR results~\cite{STAR:NPE}, the new results have greatly reduced uncertainties across the entire $\pt$ range and extend the measurements beyond central collisions. The new results are consistent with previous measurements within statistical and systematic uncertainties. 

These results are also compared to Duke (modified Langevin transport model)~\cite{ref:duke1} and PHSD (parton-hadron-string dynamics model)~\cite{ref:phsd1, ref:phsd2} model calculations shown in Fig.~\ref{Fig:Raa}. In the Duke model, heavy quarks lose energy due to quasielastic scatterings and medium-induced gluon radiation implemented using the modified Langevin equation in the medium, whose evolution is modeled according to a (2+1)-dimensional viscous hydrodynamics. Their hadronization consists of a coalescence process dominating at low $\pt$ and a fragmentation process becoming important at high $\pt$. The produced heavy-flavor hadrons are input into hadron cascade ultrarelativistic quantum molecular dynamics model~\cite{ref:urqmd} to simulate hadronic interactions. In the PHSD model, heavy quarks lose energy through elastic scattering with massive off-shell partons whose masses and widths are given by the dynamical quasiparticle model matched to the lattice QCD equation of state. Both coalescence and fragmentation processes take place during heavy quark hadronization, and the produced heavy-flavor hadrons undergo hadronic interactions described using effective field theory and taking into account resonant interactions. Both the Duke and the PHSD model calculations agree with data within uncertainties.

\begin{figure}
\centering
\includegraphics[width=.7\textwidth,page=1]{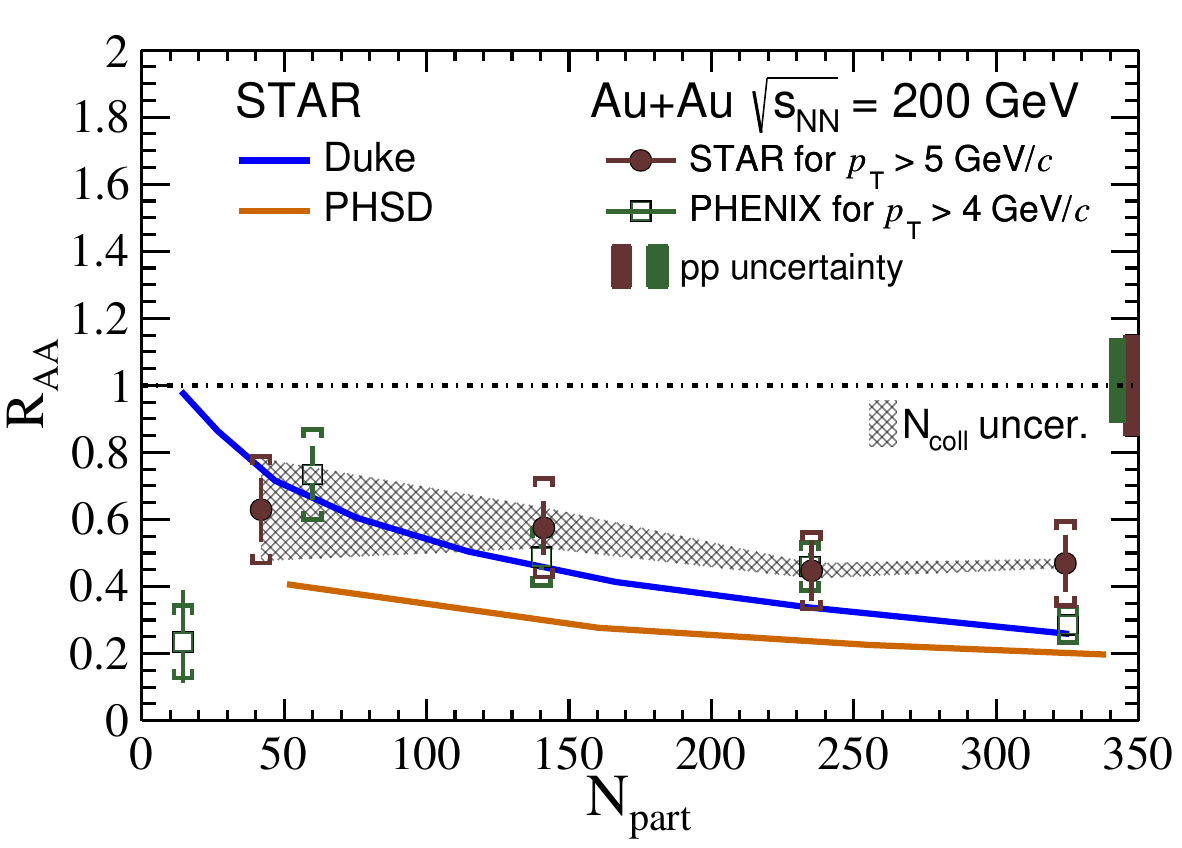}
\vspace*{-10pt}
\caption{\label{Fig:Raanpart}HFE \raa\ (red circles) as a function of \npart\ in Au+Au collisions at \sNN{200} GeV, compared with PHENIX measurements (green squares)~\cite{PHENIX:NPE}, and Duke (blue line) and PHSD (orange line) model calculations. Vertical bars and boxes around data points represent statistical and systematic uncertainties from Au+Au measurements, respectively. The gray band represents the \ncoll\ uncertainties. The boxes at unity show the global uncertainties including the total uncertainties of the $p$+$p$ reference.}
\end{figure}

The dependence of the HFE \raa\ on collision centrality, denoted as the number of participating nucleons (\npart)~\cite{D0_STAR}, for $\pt>5$ \gevc\ in Au+Au collisions at \sNN{200} GeV is shown in Fig.~\ref{Fig:Raanpart}, along with PHENIX measurement for $\pt>4$ \gevc~\cite{PHENIX:NPE}, and Duke and PHSD mode calculations. There is a hint of HFE \raa\ decreasing from peripheral to central collisions, which is in line with the expectation of stronger QGP effects in central collisions. The new results are consistent with PHENIX results within uncertainties. Both Duke and PHSD model calculations can qualitatively describe data, even though the PHSD model seems to be systematically below the central values of data.

\section{\label{sec:summary}Summary}
Measurements of HFE invariant yields and nuclear modification factors $R_{\rm AA}$ as a function of $\pt$ at mid-rapidity ($|y|<$ 0.7) for 3.5 $<\pt<$ 9 GeV/$c$ in Au+Au collisions at \sNN{200} GeV are reported. Compared to previous measurements at RHIC, the new results improve measurements of HFE suppression in the QGP with better precision above 6 GeV/$c$, and extend previous STAR measurements beyond central collisions. Approximately a factor of 2 suppression is observed in central and mid-central collisions above 3.5 \gevc, suggesting significant energy loss of heavy quarks in the hot, dense medium. Both the Duke and PHSD model calculations can qualitatively describe data within uncertainties. These results will provide an improved reference for \raa\ measurements of charm- and bottom-hadron decayed electrons in heavy-ion collisions.

\acknowledgments

We thank the RHIC Operations Group and RCF at BNL, the NERSC Center at LBNL, and the Open Science Grid consortium for providing resources and support.  This work was supported in part by the Office of Nuclear Physics within the U.S. DOE Office of Science, the U.S. National Science Foundation, National Natural Science Foundation of China, Chinese Academy of Science, the Ministry of Science and Technology of China and the Chinese Ministry of Education, the Higher Education Sprout Project by Ministry of Education at NCKU, the National Research Foundation of Korea, Czech Science Foundation and Ministry of Education, Youth and Sports of the Czech Republic, Hungarian National Research, Development and Innovation Office, New National Excellency Programme of the Hungarian Ministry of Human Capacities, Department of Atomic Energy and Department of Science and Technology of the Government of India, the National Science Centre and WUT ID-UB of Poland, the Ministry of Science, Education and Sports of the Republic of Croatia, German Bundesministerium f\"ur Bildung, Wissenschaft, Forschung and Technologie (BMBF), Helmholtz Association, Ministry of Education, Culture, Sports, Science, and Technology (MEXT) and Japan Society for the Promotion of Science (JSPS).

%
%
\bibliographystyle{JHEP}
\bibliography{paper_RS}

\appendix

\setcounter{secnumdepth}{0}

\section{The STAR Collaboration}

\bfseries\raggedright\sffamily{
M.~I.~Abdulhamid$^{4}$,
B.~E.~Aboona$^{55}$,
J.~Adam$^{15}$,
L.~Adamczyk$^{2}$,
J.~R.~Adams$^{39}$,
I.~Aggarwal$^{41}$,
M.~M.~Aggarwal$^{41}$,
Z.~Ahammed$^{62}$,
D.~M.~Anderson$^{55}$,
E.~C.~Aschenauer$^{6}$
% ...
}

\afterAuthorSpace


\normalfont\small\itshape
\begin{list}{}{%
\setlength{\leftmargin}{0.28cm}%
\setlength{\labelsep}{0pt}%
\setlength{\itemsep}{\affiliationsSep}%
\setlength{\topsep}{-\parskip}}%
\item{$^{1}$Abilene Christian University, Abilene, Texas   79699}
\item{$^{2}$AGH University of Science and Technology, FPACS, Cracow 30-059, Poland}
\item{$^{3}$Argonne National Laboratory, Argonne, Illinois 60439}
\item{$^{4}$American University in Cairo, New Cairo 11835, Egypt}
\item{$^{5}$Ball State University, Muncie, Indiana, 47306}
\item{$^{6}$Brookhaven National Laboratory, Upton, New York 11973}
\item{$^{7}$University of Calabria \& INFN-Cosenza, Rende 87036, Italy}
\item{$^{8}$University of California, Berkeley, California 94720}
\item{$^{9}$University of California, Davis, California 95616}
\item{$^{10}$University of California, Los Angeles, California 90095}
\item{$^{11}$University of California, Riverside, California 92521}
\item{$^{12}$Central China Normal University, Wuhan, Hubei 430079 }
\item{$^{13}$University of Illinois at Chicago, Chicago, Illinois 60607}
\item{$^{14}$Creighton University, Omaha, Nebraska 68178}
\item{$^{15}$Czech Technical University in Prague, FNSPE, Prague 115 19, Czech Republic}
\item{$^{16}$Technische Universit\"at Darmstadt, Darmstadt 64289, Germany}
\item{$^{17}$National Institute of Technology Durgapur, Durgapur - 713209, India}
\item{$^{18}$ELTE E\"otv\"os Lor\'and University, Budapest, Hungary H-1117}
\item{$^{19}$Frankfurt Institute for Advanced Studies FIAS, Frankfurt 60438, Germany}
\item{$^{20}$Fudan University, Shanghai, 200433 }
\item{$^{21}$University of Heidelberg, Heidelberg 69120, Germany }
\item{$^{22}$University of Houston, Houston, Texas 77204}
\item{$^{23}$Huzhou University, Huzhou, Zhejiang  313000}
\item{$^{24}$Indian Institute of Science Education and Research (IISER), Berhampur 760010 , India}
\item{$^{25}$Indian Institute of Science Education and Research (IISER) Tirupati, Tirupati 517507, India}
\item{$^{26}$Indian Institute Technology, Patna, Bihar 801106, India}
\item{$^{27}$Indiana University, Bloomington, Indiana 47408}
\item{$^{28}$Institute of Modern Physics, Chinese Academy of Sciences, Lanzhou, Gansu 730000 }
\item{$^{29}$University of Jammu, Jammu 180001, India}
\item{$^{30}$Kent State University, Kent, Ohio 44242}
\item{$^{31}$University of Kentucky, Lexington, Kentucky 40506-0055}
\item{$^{32}$Lawrence Berkeley National Laboratory, Berkeley, California 94720}
\item{$^{33}$Lehigh University, Bethlehem, Pennsylvania 18015}
\item{$^{34}$Max-Planck-Institut f\"ur Physik, Munich 80805, Germany}
\item{$^{35}$Michigan State University, East Lansing, Michigan 48824}
\item{$^{36}$National Institute of Science Education and Research, HBNI, Jatni 752050, India}
\item{$^{37}$National Cheng Kung University, Tainan 70101 }
\item{$^{38}$Nuclear Physics Institute of the CAS, Rez 250 68, Czech Republic}
\item{$^{39}$The Ohio State University, Columbus, Ohio 43210}
\item{$^{40}$Institute of Nuclear Physics PAN, Cracow 31-342, Poland}
\item{$^{41}$Panjab University, Chandigarh 160014, India}
\item{$^{42}$Purdue University, West Lafayette, Indiana 47907}
\item{$^{43}$Rice University, Houston, Texas 77251}
\item{$^{44}$Rutgers University, Piscataway, New Jersey 08854}
\item{$^{45}$Universidade de S\~ao Paulo, S\~ao Paulo, Brazil 05314-970}
\item{$^{46}$University of Science and Technology of China, Hefei, Anhui 230026}
\item{$^{47}$South China Normal University, Guangzhou, Guangdong 510631}
\item{$^{48}$Sejong University, Seoul, 05006, South Korea}
\item{$^{49}$Shandong University, Qingdao, Shandong 266237}
\item{$^{50}$Shanghai Institute of Applied Physics, Chinese Academy of Sciences, Shanghai 201800}
\item{$^{51}$Southern Connecticut State University, New Haven, Connecticut 06515}
\item{$^{52}$State University of New York, Stony Brook, New York 11794}
\item{$^{53}$Instituto de Alta Investigaci\'on, Universidad de Tarapac\'a, Arica 1000000, Chile}
\item{$^{54}$Temple University, Philadelphia, Pennsylvania 19122}
\item{$^{55}$Texas A\&M University, College Station, Texas 77843}
\item{$^{56}$University of Texas, Austin, Texas 78712}
\item{$^{57}$Tsinghua University, Beijing 100084}
\item{$^{58}$University of Tsukuba, Tsukuba, Ibaraki 305-8571, Japan}
\item{$^{59}$University of Chinese Academy of Sciences, Beijing, 101408}
\item{$^{60}$United States Naval Academy, Annapolis, Maryland 21402}
\item{$^{61}$Valparaiso University, Valparaiso, Indiana 46383}
\item{$^{62}$Variable Energy Cyclotron Centre, Kolkata 700064, India}
\item{$^{63}$Warsaw University of Technology, Warsaw 00-661, Poland}
\item{$^{64}$Wayne State University, Detroit, Michigan 48201}
\item{$^{65}$Yale University, New Haven, Connecticut 06520}
\end{list}

\end{document}